\documentclass[preprint,pre,groupedaddress]{revtex4}

\usepackage{amsmath}
\usepackage{amsfonts}
\usepackage{amssymb}
\usepackage{graphicx}
\usepackage{color}
\usepackage{epstopdf}
\usepackage{xcolor,colortbl}
\usepackage[thinlines]{easytable}
\usepackage{caption}
\usepackage{placeins}

\definecolor{gray}{rgb}{0.5,0.5,0.5}
\definecolor{purple}{rgb}{0.9,0.2,0.6}
\newcommand{\SI}[1]{{{#1}}}

 \usepackage{ulem}

\begin{document}


\textbf{\large{Atomic rheology of gold nanojunctions}}

{Jean Comtet}, {Antoine Lain\'e}, {Antoine Nigu\`es}, {Lyd\'{e}ric Bocquet}, {Alessandro Siria}*

\textit{Laboratoire de Physique Statistique de l'Ecole Normale Sup\'erieure, UMR CNRS 8550, PSL Research University, 24 Rue Lhomond 75005 Paris, France}

* alessandro.siria@lps.ens.fr
\newline

{\bf
Despite extensive investigations of dissipation and  deformation processes in micro- and nano- sized metallic samples \cite{Kraft2010, Wu2005,Yu2010,Chen2006a, Rubio1996,Kuipers1993, Sun2014}, the mechanisms at play during deformation of systems with ultimate, molecular size remain elusive. 
While metallic nanojunctions, obtained by stretching metallic wires down to the atomic level are a system of choice to explore atomic scale contacts \cite{Rubio1996,Yanson1998,Ohnishi1998,Shiota2008, Agrait2003}, it has not been possible up to now to extract the full equilibrium and out of equilibrium rheological flow properties of matter at such scales. Here, by using a quartz-tuning fork based Atomic Force Microscope (TF-AFM), we combine electrical and rheological measurement on angstr\"om-size gold junctions to study the non linear rheology of this model atomic system. By submitting the junction to increasing sub-nanometric deformations we uncover a transition from a purely elastic regime to a plastic, and eventually to a viscous-like fluidized regime, akin to the rheology of soft yielding materials \cite{Mason1996,Larson1999,Bocquet2009}, though orders of magnitude difference in length scale. The fluidized state furthermore highlights  capillary attraction, as expected for liquid capillary bridges. This shear fluidization cannot be captured by classical models of friction between atomic planes \cite{Frenkel1926a,Zaloj1998}, pointing to unexpected dissipative behavior of defect-free metallic junctions at the ultimate scales. Atomic rheology is therefore a powerful tool to probe the structural reorganization of atomic contacts.
}
\newline

Solid metallic materials in the micron range and below have been shown to exhibit drastically different mechanical behavior as compared to their macroscopic counterparts. Such size effects take origin in the decreasing density of defect-mediated plastic events \cite{Kraft2010}, such as occurring during dislocation gliding \cite{Wu2005} or twinning \cite{Yu2010}, as well as an increasing surface to volume ratio \cite{Chen2006a}. However, extending measurements of the mechanical response to the $\sim$10 nm scale and below has been a challenging task, leading to apparently contradicting measurement reporting both very large yield stress \cite{Rubio1996, Wu2005} and pseudo-elastic deformation 
\cite{Kuipers1993, Sun2014}. The understanding of plastic flow and, in more general terms, of dissipation in systems of molecular or atomic sizes has so far remained elusive, despite its fundamental interest and broad applications ranging from the understanding of shape stability in nanoelectronics \cite{Lu2007}, fundamental dissipation channels in nanomechanical resonators \cite{Guttinger2017}, as well as understanding of the role played by nanocontacts in macroscopic friction and adhesion \cite{Persson2005}

Gold nanojunctions, obtained by stretching metallic wires \ down to the atomic level, have been extensively studied for their electronic properties in the past two decades \cite{Yanson1998,Ohnishi1998}, their molecular dimensions leading to quantized electrical conductance. Here we divert from their standard application and, by means of TF-AFM, we explore the full equilibrium and out-of-equilibrium rheological flow behaviors of atomic junctions, giving unprecedented insights into the dissipative mechanisms at play in those systems of few atoms.

\begin{figure}[!htb]
\centering
\includegraphics[width=\columnwidth]{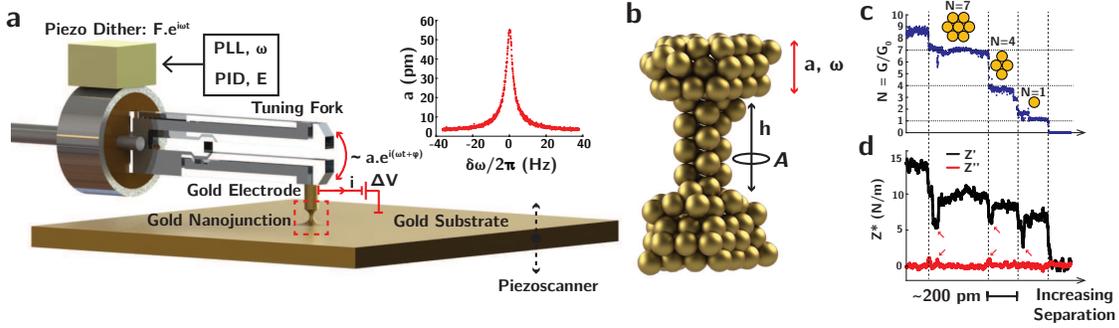}
\caption{{\textbf{FIGURE 1: Experimental Set-Up} (\textbf{a}) Schematic of the experimental set-up. {A quantum point contact, consisting of a nanojunction of gold} (red dashed box) is formed between a gold electrode, attached to the tuning fork and a gold substrate. A bias $\Delta V$ [V] is applied between the electrode and the substrate, allowing measurements of the junction conductance $G$ [S]. Gold substrate position is controlled with a piezoscanner. A piezo dither excites the tuning fork with an oscillatory force $F^* = F\cdot e^{i\omega t}$, leading to the oscillation $a^* = a\cdot e^{i (\omega t+\phi)}$ of the gold electrode and tuning fork. {Two feedback loops tune the excitation frequency $\omega$ [Hz]  (Phase-Locked Loop PLL) and excitation force $F$ [N] (PID) to systematically excite the tuning fork at resonance and maintain a set oscillation amplitude}, allowing direct measurement of the gold nanojunction viscoelastic impedance $Z^* = F^*/a^*$ (see {\SI{Methods section 'Measurement of viscoelastic properties'}}). {Inset: resonance curve of the free tuning fork in air.} (\textbf{b}) Schematic representation of the idealized junction geometry, for $\mathcal{N} = G/G_0 \approx 4$. The junction is assumed to have a rod-like shape with height $h$ and surface area $\mathcal{A} \approx \mathcal{N} \pi d_0^2/4$. (\textbf{c}) Typical conductance trace for increasing separation between the two surfaces. Conductance varies stepwise, in multiples of $G_0$. (\textbf{d}) Simultaneous measurement of the conservative (Z', black) and dissipative (Z'', red) part of the mechanical impedance $Z^*$ of the gold junction, here in the low amplitude (elastic) regime.} Red arrows indicate transient decrease in stiffness $Z'$ and increase in dissipation $Z''$ interpreted as a signature of plastic reorganization of the junction.}
\end{figure}

During a typical experiment, a gold electrode, glued on one prong of TF-AFM, is indented on a gold-coated surface, creating a gold junction (Fig. 1a-b). As observed in Fig. 1c, as the junction thins down due to the increasing distance between the electrode and the substrate, we observe a step-wise variation in the conductance $G$ of the junction, which varies approximately in multiples of $G_0 = 2 e^2/h \approx 77~\mu$S. Quantization of the conductance occurs as the molecular lateral size of the junction leads to ballistic electronic transport, for which conductance $G$ is simply proportional to the number $\mathcal{N}$ of conductance channels, aka to the number of atoms in the cross-section, with  $G~\approx~\mathcal{N}\cdot G_0$. This quantization allows a direct readout of the transverse dimension of the junction at the single atom level. Fig. 1c shows the successive thinning of the junction from $\mathcal{N} =7$, down to $\mathcal{N} = 1$ atoms.

To simultaneously probe the mechanical properties of the junction, we further excite the TF-AFM via a piezo dither, with a periodic force $F^* = F \exp(i \omega t)$ [N], leading to subnanometric oscillations $a^* = a \exp(i \omega t + \phi)$ of the tuning fork and upper gold electrode. The viscoelastic behavior of the gold nanojunction can be characterized by the complex impedance $Z^* = F^*/a^* = F/a \cdot \exp(-i\phi)$ [N/m], where the real part $Z'=\Re[Z^*]$ and imaginary part $Z'' = \Im[Z^*]$ characterize respectively the conservative (elastic) and dissipative response of the junction {(See Fig. 1a and \SI{Methods section 'Measurement of viscoelastic properties'})}.

We show in Fig.~1d, the variations of $Z'$ and $Z''$ measured with a fixed oscillation amplitude $a = 70$~pm, along with the variations of junction conductance during elongation (Fig.~1c). On each plateau in conductance, the stiffness $Z'$ is approximately constant, indicating a constant mechanical structure of the junction, while the dissipative modulus $Z''$ is vanishingly small pointing out to the absence of intrinsic dissipation in the structure for such deformations. At each change in conductance, we observe transient decrease in stiffness $Z'$ and slight increase in dissipation $Z''$ (red arrows, Fig. 1d), which can be interpreted as signatures of plastic reorganization during the transformation between states of different contact size. Those measurements are consistent with previous measurements on metallic junctions, reporting that elongation is caused by successive elastic and yielding events, concomitant with changes in the electrical conductance \cite{Rubio1996, Agrait2003, Shiota2008, Marszalek2000}.

\begin{figure}[!htb]
\centering
\includegraphics[width=0.5\columnwidth]{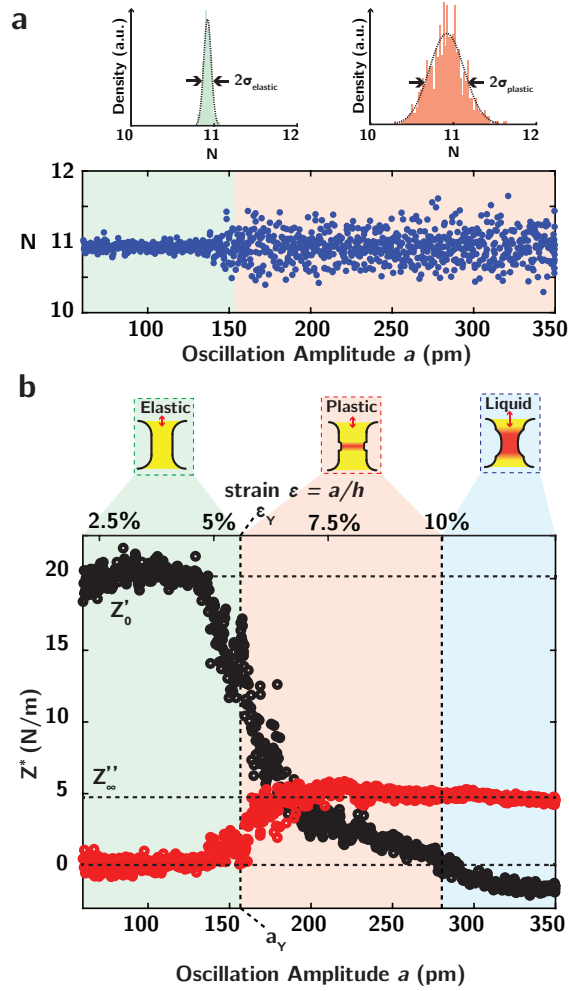}
\caption{{\textbf{FIGURE 2: Atomic Rheology.} \textbf{(a)}~Dimensionless conductance trace $\mathcal{N} = G/G_0$ as a function of the oscillation amplitude of the TF-AFM for sampling rate of 50 Hz. At the transition between the elastic and plastic regimes of deformation, the current noise increases, while retaining a fixed mean value. Histograms show current distribution for the elastic (green) and plastic (red) regimes and the corresponding distribution width $\sigma$. \textbf{(b)} Measurement of the viscoelastic properties of the junction with a fixed cross-sectional area.  Variation of storage ($Z'$, black) and loss ($Z''$, red) impedance of the junction as a function of oscillation amplitude $a$ of the TF-AFM. With increasing oscillation amplitude, we observe a successive transition from elastic (green, constant $Z' = Z'_0$, $Z'' = 0$) to plastic (red, decrease in $Z'$, increase an plateau in $Z'' = Z''_\infty$) up to a liquid-like regime showing capillary adhesion (blue, $Z'<0$) }}
\end{figure}
Going beyond this simple picture requires a distinct strategy to probe the full out-of-equilibrium rheological flow properties and dissipative mechanisms at play in the junction. Our methodology consists in maintaining a fixed number of atoms in the cross-section through a feed-back control of the junction conductance $G$, and submitting the junction to increasing shear to probe its viscoelastic flow properties under a wide range of shear rate and shear stress (with shear amplitude ranging from 20 pm up to 1 nm, see \SI{Extended Data Fig. 5}). This approach is similar in spirit, although at radically different length scales, to the exploration of the mechanical response of foams and emulsions \cite{Mason1996,Larson1999}, and allows for a quantitative measurement of the rheological response at the atomic scale.

As shown in Fig. 2a for a contact number of $\mathcal{N} = 11$ atoms, the fluctuations in the dimensionless conductance trace $\mathcal{N}=G/G_0$ increase at large oscillation amplitude, but the mean contact conductance and thus the averaged junction geometry remains fixed. In Fig. 2b, simultaneous measurements of the elastic ($Z' = \Re(Z^*)$, black) and dissipative ($Z'' = \Im(Z^*)$, red) part of the mechanical impedance $Z^*$ [N.m$^{-1}$] allows to extract the full linear and non-linear rheological flow properties of the junction. Three consecutive regimes are highlighted as a function of the oscillation amplitude, from elastic to plastic, to liquid-like behavior. Remarkably, in spite of orders of magnitude of difference in scales,  this overall phenomenology echoes similar trends for plastic flows in soft yielding materials \cite{Mason1996,Larson1999}.

At low oscillation amplitude, the gold junction is unperturbed, and we observe a purely elastic response, characterized by the absence of dissipation in the junction ($Z'' \simeq 0$) and a finite positive stiffness $Z'_0>0$ (Fig. 2b, green region). 
Accordingly, current fluctuations are found to be small (Fig. 2a, green histogram with spectral density $S_\mathcal{N} =  0.01~\text{Hz}^{-1/2}$, defined in terms of the distribution of $\mathcal{N}$ as $S_\mathcal{N}=\sigma/\sqrt{\nu}$, with $\nu$ [Hz] the sampling frequency). The constant elastic stiffness $Z'_0$ in the elastic regime allows a coarse characterization of the contact height $h$ (Fig. 1b) which is found typically in the range $h\approx 1-6$ nm \SI{(See Methods section 'Geometry of the junction')}.

As shown in Fig. 2b, increasing the oscillation amplitude $a$ for a fixed contact size $\mathcal{N}$, we evidence an abrupt decrease in the stiffness $Z'$  and a corresponding increase in the dissipative response $Z''$ (going from green to red zone in Fig.~2b; note that rheological curves are completely reversible upon decrease of the shear rate, see \SI{Extended Data Fig. 6}).
 This dramatic change points to a dissipative reorganisation of the junction under shear, which is further evidenced by the increase in current fluctuations (Fig. 2a, red histogram with spectral density $S_\mathcal{N} = 0.04~\text{Hz}^{-1/2}$). 
This mechanical response is in striking analogy with the (non-linear) rheology of soft yielding materials, such as foams and emulsions \cite{Mason1996,Larson1999} 
 where it is  a signature of the onset of yielding and plasticity.
Here this regime occurs above a threshold oscillation amplitude $a_\text{Y}$, which we define  as the point where $Z'_0$ decreases and $Z''_\infty$ increases to half their asymptotic values ($Z'(a_Y) \approx Z'_0/2$, roughly concomitant to $Z''(a_Y) \approx Z''_\infty/2$, {alternative definitions giving similar results}).

\begin{figure}[!htb]
\centering
\includegraphics[width=0.9\columnwidth]{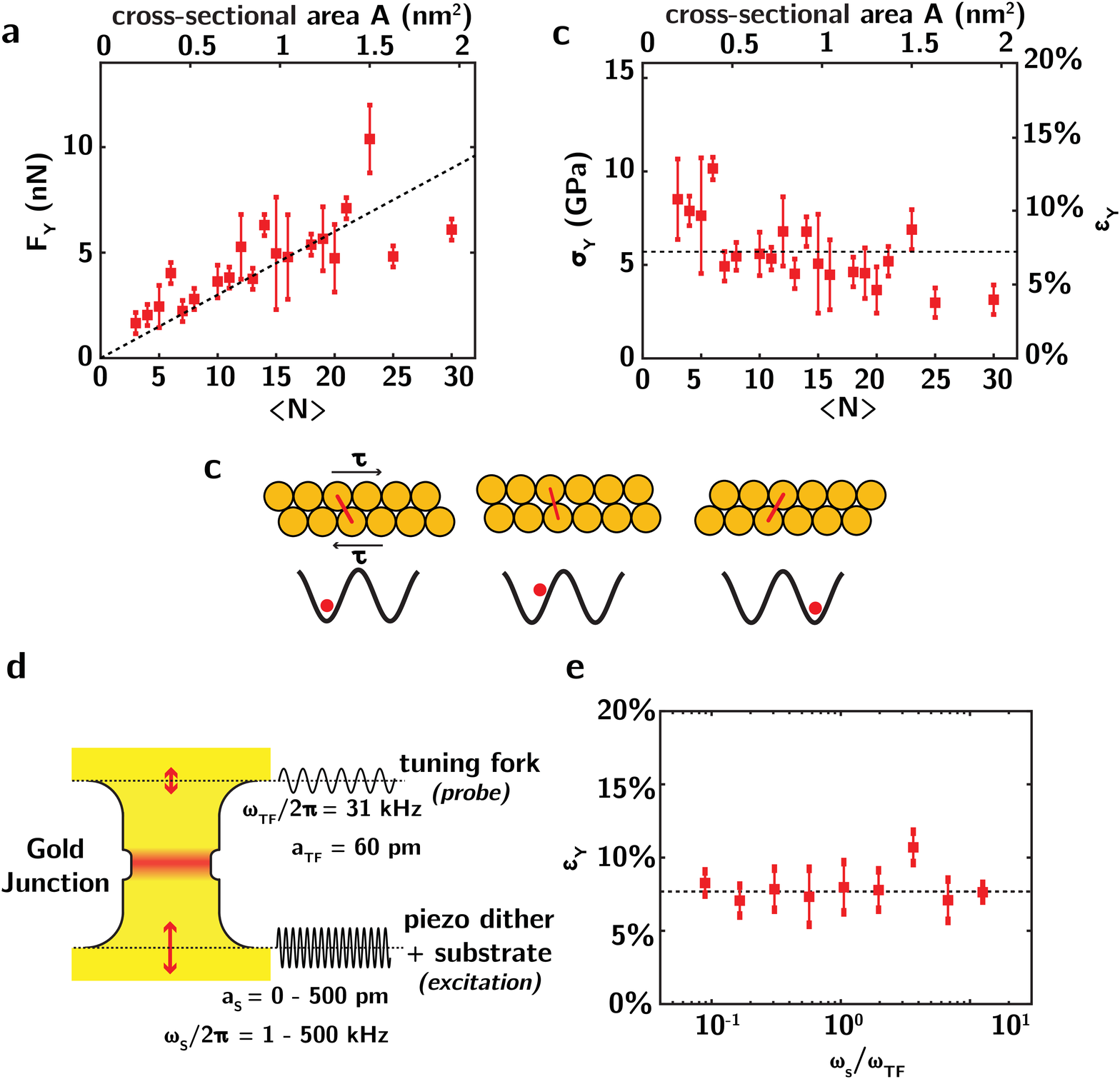}
\caption{{\textbf{FIGURE 3: Yielding Threshold.} (\textbf{a}) Yield force $F_\text{Y}$,  (\textbf{b}) yield stress $\sigma_\text{Y}$ (left axis) and yield strain $\epsilon_\text{Y}$ (right axis) as a function of cross-sectional atom number $\mathcal{N}$ (and surface area $\mathcal{A}$ of the junction, see upper axis).  (\textbf{c}) Slip in a perfect crystal between two atomic planes under shear stress $\tau$. (\textbf{d}) Schematic of the set-up. An additional piezo dither is placed below the substrate, allowing to shear the junction at an additional frequency $\omega_\text{s}/2\pi$,  while the junction properties are simultaneously probed by the TF-AFM  (see \SI{Methods section 'Effect of excitation frequency on the plastic transition'}). (\textbf{e}) yield strain $\epsilon_\text{Y}$ as a function of the additional excitation frequency $\omega_\text{s}$ (respective to the TF-AFM frequency $\omega_{\rm TF}$) for $\mathcal{N} = 15$. Error bars represent standard deviation and are taken to 10 \% for single valued points.}}
\end{figure}

Using the quantized variation of the junction conductance with lateral number of atoms ($G=\mathcal{N}\cdot G_0$), we now vary -- atom by atom -- the lateral  size of the junction, from sizes of $\mathcal{N} \approx 30$ atoms down to $\mathcal{N} \approx 3$ atoms, corresponding to an equivalent cross-sectionnal area $\mathcal{A} \approx \mathcal{N} \pi d_\text{gold}^2/4 \approx 2-0.2$ nm$^2$ \SI{(Extended data Fig. 5)}.
As shown in Fig.~3a, the yield force, defined as $F_Y = Z'_0 \cdot a_\text{Y}$, is roughly proportional to the mean number $\mathcal{N}$ of atoms in a cross-section (lower axis) or equivalently the cross-sectional area $\mathcal{A}$ (upper axis).

The corresponding yield stress $\sigma_\text{Y} = F_\text{Y}/\mathcal{A}$ is accordingly found to be roughly independent of contact area and of the order of $5$ GPa (Fig.~3b, left axis), while the yield strain $\epsilon_\text{Y}=a_\text{Y}/h = \sigma_\text{Y}/E_\text{gold}$ is found of the order 5-10\% (Fig. 3b, right axis).

Importantly, those values are much larger than in macroscopic gold samples, for which $\sigma_\text{Y}$ ranges from 55 to 200 MPa \cite{Espinosa2004} or in single crystal gold nano-pillars, where $\sigma_\text{Y}$ is of the order of 500 MPa \cite{Lee2009}. Such high values of yield stress and yield strain are consistent with the expected absence of defects in atomic size junctions. 

As a first modeling approach, one may compare the values of the experimental yield stress to a Frenkel-type of estimate
 \cite{Frenkel1926a, Agrait2003,Zaloj1998}, based on the slippage of a perfect crystal {(Fig. 3c)}.
For slip along the \{111\} plane of a fcc crystal, the resolved shear stress is given by $\tau_\text{max} \approx  G/9$.
Taking gold shear modulus $G = 27$ GPa and a Schmid factor $m \approx 0.5$, we find a yield stress $\sigma_\text{Y} \approx \tau_\text{max}/m \approx 6$~GPa, in good agreement with our measurements (Fig. 3b). This confirms that the junction yields merely as a defect free ordered crystal, which is to be expected for such size.

 An important question is to understand the respective influence of shear stress and shear rate on the observed plastic transition. To probe this dependence, we submit the junction to additional oscillatory strains at various frequencies, as sketched in {Fig.~3d}. Plastic flow is here {\it induced} by this external shear while junction properties are simultaneously {\it probed} with the TF-AFM oscillating at constant amplitude (\SI{see Methods section 'Effect of excitation frequency on the plastic transition' and Extended data Fig. 3}). Such experiments is similar in spirit to superposition rheometry  \cite{Vermant1997}. As shown in {Fig.~3e}, these complementary measurements demonstrate that the yield strain for the plastic transition is basically independent of the frequency of the oscillatory strain. The plastic transition is accordingly driven by a critical strain, and not a critical strain rate, as expected for a system with no intrinsic timescales.

Going more into details of the $\sigma_Y(\mathcal{N})$ and $\epsilon_Y(\mathcal{N})$ curves, one may remark that 
the yield strain and yield stress appear to exhibit local maxima for a number of specific values of the conductance channel number $\mathcal{N}$, typically $\mathcal{N}\approx 6, 12, 14, 23$. Interestingly these values are close to the (so-called) ``magical'' numbers for the more stable gold wires predicted theoretically for $\mathcal{N}\approx 7, 11, 14-15, 23$ in \cite{Tosatti2001} and observed in Transmission Electron Microscope \cite{Kondo2000}. This would suggest that the more stable gold wires presents a larger yield strain to enter in a plastic regime. This observation however deserves a dedicated exploration which we leave for future work. 

We now turn to the regime of large deformations (Fig. 2b, red zone). Surprisingly, the dissipative modulus exhibits a plateau in this regime and thus becomes independent of the imposed deformation $a$, while the junction stiffness decreases steadily (Fig.~2b). The corresponding dissipative force is accordingly expected to increase linearly with deformation, as $F_\text{D} \approx Z''_\infty \times a$, suggesting a ``viscous-like'' dissipation, for which dissipation is proportional to the imposed velocity with a linear response between driving and dissipation.

One would accordingly expect $F_\text{D} = \eta \dot \gamma \mathcal{A}$, with $\dot \gamma \sim a \omega/h$ $\sim 10^4$ s$^{-1}$ the typical shear-rate, $\mathcal{A}$ the contact area and $\eta$ a material viscosity. 

To further assess these views,  we plot in Fig.~4a the friction coefficient $F_\text{D}/\dot\gamma$ -- defined as  $F_\text{D}/\dot\gamma=Z''_\infty\cdot h/\omega$ -- as a function of the contact lateral size $\mathcal{N}$ (lower axis) or  cross section $\mathcal{A}$ (upper axis). 
This plot confirms that the dissipative force is proportional to the contact area, allowing us to infer a viscosity $\eta$ (Fig. 4b), which is found to be roughly constant and of the order of $\eta \sim 7 \cdot10^4$~Pa.s at the TF-AFM excitation frequency $\omega_\text{TF}/2\pi \approx 31$~kHz. 

This viscous-like regime is completely unexpected for defect-free crystalline systems (Fig. 2c), for which plastic flow should occur at {\it constant stress} \cite{Zaloj1998,Frenkel1926a} {(see \SI{Methods section 'Comparison with Prandtl-Tomlinson model' and Extended Data Fig. 7}).

To get further insights in this viscous-like behavior, one may define a Maxwell time-scale $\tau_{\rm M}$ characterizing relaxation of the system as $\eta=G_{\rm gold} \cdot \tau_{\rm M}$ with $G_{\rm gold} =27$ GPa the shear modulus of gold \cite{Larson1999}. The corresponding value $\tau_{\rm M} \sim 3~\mu$s is huge as compared to microscopic time scales (typically in the picosecond range), and very close to the excitation time scale $\sim 1/\omega_\text{TF}=5~\mu$s. This therefore suggests that the excitation does fix the relaxation time-scale of the junction under strong deformation, in direct line with the behavior of yielding materials -- emulsions, foams or granular materials -- where the fluidity (inverse viscosity) is fixed by the excitation time-scale itself \cite{Bocquet2009,Eisenmann2010, Dijksman2011}.

\begin{figure}[!htb]
\centering
\includegraphics[width=0.9\columnwidth]{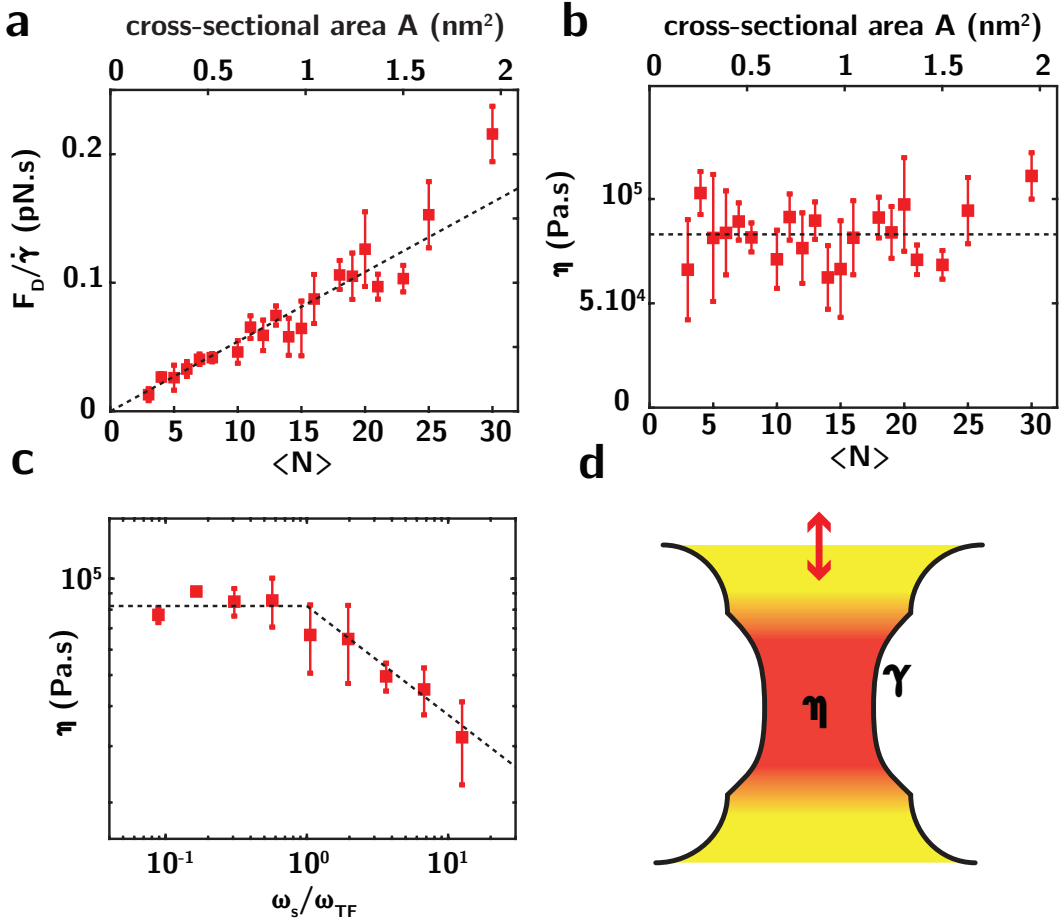}
\caption{{\textbf{FIGURE 4: Liquid-like behavior.} (\textbf{a})  Dissipative hydrodynamic force $F_\text{D}/\dot\gamma$, with $\dot\gamma=a\omega/h$ as a function of cross-sectional atom number $\mathcal{N}$ (and surface area of the junction, see upper axis). (\textbf{b}) Viscosity $\eta$ as a function of $\mathcal{N}$. (\textbf{c}) Measured viscosity as a function of oscillation frequency $\omega_\text{s}/\omega_\text{TF}$ for $\mathcal{N} = 15$.  Error bars are standard deviation and are taken to 10 \% for single valued points.  {(\textbf{d}) Liquid-like response of the junction at large deformation, with signatures of both viscous-like dissipation (viscosity $\eta$) and capillary adhesion (surface tension $\gamma$).}}}
\end{figure}

To assess this dependence of the viscous behavior on excitation frequency, we measure as shown in Fig. 3c the viscosity $\eta$ of the liquefied gold junction {\it induced} by an additional excitation frequency $\omega_\text{s}$ (see Fig. 3c, \SI{Methods section 'Effect of excitation frequency on the plastic transition' and Extended Data Fig. 2}). As shown in Fig. 4c, the measured viscosity is found to be roughly independent of excitation frequency for $\omega_\text{s}/\omega_\text{TF}<1$, as the response of the junction is accordingly fixed by the lowest excitation time $1/\omega_\text{TF}$. For larger excitation frequency $\omega_\text{s}/\omega_\text{TF}>1$, we observe a thinning behavior with $\eta \sim \eta_\text{TF} (\omega_\text{s}/\omega_\text{TF})^{-\alpha}$ with $\alpha \approx 0.28$ and $\eta_\text{TF} \approx 7 \cdot10^4$~Pa.s. This thinning behavior further highlights the dominant role of excitation frequency in the liquid-like dissipative behavior of the junction.

Our experiments thus point to unexpected dissipation channels during the deformation of the junction under large strain. Keeping in mind the yielding process in macroscopic soft materials \cite{Bocquet2009,Larson1999}, such fluidization under stress might be due to the collective atomic reconstruction process, potentially favored by the rapid surface diffusion of gold atoms \cite{Kuipers1993, Sun2014, Gosvami2011, Merkle2008}. 
Such additional dissipation channels, completely unexpected for dislocation-free ordered crystalline systems open exciting perspective for the fundamental modeling of dissipative processes at the atomic scale. Finally, the slight deviation from linearity observed for $N>20$ in Figs. 3a and 4a suggests a possible transition to more traditional dislocation-based mechanisms as sample volumes becomes sufficiently large.

Surprisingly, the liquid-like character of the gold junction is also recovered in the conservative elastic response in the form of a negative stiffness under large strain (Fig. 2b, blue zone). 

This regime is therefore associated  with an attractive adhesive response of the junction, which is reminiscent of capillary adhesion of macroscopic capillary bridges \cite{Crassous1997} (Fig. 4d). This additional signature of the liquid-like behavior of the junction at large oscillation amplitudes, whose conservative mechanical response becomes dominated by surface effects is further confirmed by additional force spectroscopy measurements in vacuum, showing "jump-to-contact" of the liquified gold (\SI{Methods section 'Capillary attraction at large oscillation amplitude' and Extended Data Fig. 3}). Effect of local junction heating on the observed adhesive behavior can be discarded by order of magnitude estimates (\SI{Methods section 'Energy balance for the shear induced fluidization of the junction' and Extended Data Fig. 4}). 

Those observations allow us to estimate the surface stress of the liquefied gold meniscus. Using the expression for the adhesive force induced by a perfectly wetting liquid between two spherical contacts, one get a stiffness $Z' \approx - 2 \pi \gamma \left({R}/{h} \right)$ {(Fig. 4d)} \cite{Crassous1997}. Identifying the radius $R$ with that of the cross-sectional area of the bridge (Fig. 2b), we find for the experiment of Fig. 2, $R/h \sim 0.17$ and $Z' \approx -2$ N.m$^{-1}$, leading to the estimate of the surface stress as $\gamma \approx 2$~N.m$^{-1}$. This value is in fair agreement with the expected value of $1$~N.m$^{-1}$ for the surface tension of liquid gold \cite{Kaufman1965}. 

Altogether our study allowed for the first time the investigation of the full equilibrium and out-of-equilibrium atomic rheology of gold junction with single atom resolution. Our measurements highlight a counter-intuitive fluidization of the gold junction under large imposed deformation, akin to soft matter yielding materials. Ultimately, fluidization leads to the complete liquid-like response of the junction under large deformation, with signatures of both viscous-like dissipation and capillary-like adhesion. Such viscous-like response is not captured by standard mechanical models of crystalline interfaces and would require a full description of the atomic reconstruction under the imposed stress as a supplementary ingredient. This opens new perspectives in the modeling of defect-free metallic materials and echoes recent simulations of bulk metal plasticity at large strains \cite{Zepeda2017}.
Finally, those experiments could find interesting applications in the context of ultrasonic and cold welding \cite{Lu2010}, as well as a better understanding of macroscopic friction based on individual atomic contact mechanics \cite{Persson2005}. 
%
%
%
%
%

\end{document}